\documentclass[aps,twocolumn,showpacs,amsmath,amssymb,prl]{revtex4}
\usepackage{graphicx}
\usepackage{dcolumn}
\usepackage{bm}

\begin{document}

\title{Probing the mechanical unzipping of DNA}

\author{N. K. Voulgarakis}
\author{A. Redondo}
\author{A. R. Bishop}
\author{K. \O. Rasmussen}

\affiliation{
Theoretical Division and Center for Nonlinear Studies, 
Los Alamos National Laboratory, Los Alamos, New Mexico 87545, USA
}

\date{\today}

\begin{abstract}
A study of the micromechanical unzipping of DNA in the framework of the Peyrard-Bishop-Dauxois model is presented. We introduce 
a Monte Carlo technique that allows accurate determination of the dependence of the unzipping forces on unzipping speed and temperature.
Our findings agree quantitatively with experimental results for homogeneous DNA, and for  $\lambda$-phage DNA we reproduce the 
recently obtained experimental force-temperature phase diagram.  Finally, we argue that there may be fundamental differences between  {\em in vivo} and {\em in vitro} DNA unzipping.
\end{abstract}


\maketitle

Separation of double-stranded DNA (dsDNA) into single-stranded
DNA (ssDNA) is fundamental to DNA replication and other important
intra-cellular processes in living organisms. In equilibrium,
DNA will denaturate when the free energy of the separated ssDNA
is less than that of the dsDNA. Because of the larger entropy of the flexible 
single-strand, this can most easily be achieved by 
increasing the temperature
of the sample until the DNA melts, somewhat above body temperature. In living organisms,
however, DNA separation is not only thermally driven, but also
enzymes and other proteins may force the two strands apart. 

Recent advances in single-molecule force spectroscopy and dynamical force spectroscopy (DFS)
 has made possible the systematic investigation of force-induced separation of
dsDNA at room temperature where dsDNA is thermally stable in the
absence of an applied force \cite{ForceRev}. Although, these studies have significantly enhanced 
the understanding of the mechanical aspects of DNA replication and 
transcription {\em in vivo}, it is also imperative for further developments of technologies, such as 
polymerase chain reaction and DNA chips, to understand the relation between 
thermal denaturation and force-induced separation. An initial step in this direction was taken by 
Danilowicz {\em et al.} \cite{pd_exp} who published an experimentally determined phase diagram for the denaturation temperature 
as a function of the applied force. This study showed, as theoretically predicted \cite{nelson}, that the force required to unzip 
the DNA decreases with increasing temperature. However the applied theoretical framework \cite{pd_exp} does not 
capture all details in the entire temperature range. 

The mechanical unzipping of DNA has also been a subject of several theoretical studies, which have often
concentrated on macroscopical aspects by investigating thermodynamic equilibrium conditions \cite{heslot}. 
Simulating realistic dynamics is unfeasible since the time scales reachable in molecular dynamics simulations are 
orders of magnitude smaller than in experiments \cite{PeyrardForce}.  In this work we present a simple and efficient numerical Monte Carlo (MC) approach, to 
describe the unzipping process macroscopically as well as semi-microscopically. In particular, we provide a theoretical underpinning for the 
experimental force-temperature phase diagram recently published by Danilowicz {\em et al.} \cite{pd_exp}.

For this purpose, we use a simple one-dimensional model of DNA proposed by Peyrard-Bishop-Dauxois (PBD) \cite{PB}. 
This model has been demonstrated \cite{italians, kimtra, ares} to describe the thermally generated large amplitude local fluctuations 
quite accurately, an aspect which is generally ignored in the thermodynamic models.
We will show that this model successfully describes many of the micro-mechanical unzipping properties of DNA and provides significant 
insights into the  physical phenomena governing unzipping.

In a typical unzipping experiment, one strand of the DNA molecule is tethered, at the terminating base, to a fixed
surface while the other strand is connected through a polymeric linker to a force probe, such as
a laser trap or an atomic force microscope (AFM) cantilever. 
For the sake of simplicity, the force probe
is usually treated as a linear spring whose elastic properties can be determined through the system calibration.
The force required to keep the molecule extended at a given distance is determined by 
measuring the deflection of the force probe.
\begin{figure}[h]
\vspace*{-0.cm}
\includegraphics[width=6.5cm,angle=0]{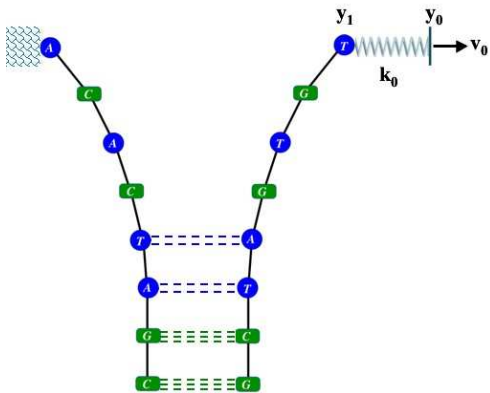}
\caption{\label{fig:model}
A schematic representation of the PBD model in a dynamic force spectroscopy experiment.}
\vspace*{-.0cm}
\end{figure}
By pulling the molecule apart at a constant speed,
the force-extension curves of the system can be obtained.
It should be mentioned, however, that the force determined by such techniques 
cannot be consider a characteristic quantity of the DNA, because it strongly depends on the elasticity of 
the experimental setup, as well as on the pulling speed. However, by performing the experiment at a wide 
range of pulling speeds and temperatures, useful
information regarding the energy landscape of the DNA unzipping process can be accumulated.

Similarly to this typical experimental setup, we use the model schematically represented in Fig. \ref{fig:model}.
In this model there is only one degree of freedom describing the relative displacement of each base pair (bp) from its equilibrium position. 
The hydrogen bonds that hold the two strands together are approximated by a Morse potential (dashed lines), while the stacking interaction 
between successive base-pairs is described by nonlinear springs (solid black line).
The model's first base to the left is fixed, while the complementary base is attached to a linear spring (force probe) of stiffness $k_0$ which moves
at a constant speed $v_0$. The force probed by the spring is therefore given by Hook's law, $V'_{pull}=k_0(y_0-y_1)$, where $y_1$ and $y_0$ are the displacements (see Fig. \ref{fig:model})
of the first base-pair and the spring's opposite end point, respectively. 
Specifically, the potential energy of the model is :
\begin{align}
\label{eq:Hamil}
V=&\sum_n \Big[D_n(e^{-a_ny_n}-1)^2 + \nonumber\\
  &\frac{k}{2}(1+\rho e^{-b (y_n+y_{n-1})})(y_n-y_{n-1})^2 \Big]+V_{pull}.
\end{align}
The PBD model parameters are those determined by Campa and  Giansanti \cite{italians}.
~To compute the force applied during the unzipping process, we performed Monte Carlo simulations: 
The force probe is moved to the right ($y_0 \rightarrow y_0 +\Delta y$) by the distance $\Delta y$ after which
$N$ Monte Carlo steps are performed in order to compute the average displacement,  $\langle y_1 \rangle $ of the first base pair. The force  will then be  given by $F=k_0(y_0-\langle y_1 \rangle)$. 
The more MC steps used for sampling, the closer to equilibrium the system is before the next $\Delta y$ move occurs.
Thus, in this Monte Carlo framework, the pulling speed can be defined as $v_0=\Delta y/N$. 
However, it should be emphasized  that no direct comparison to the real time can be made as this depends on the detailed
implementation of the Monte Carlo method. Here, the unit of pulling speed is $V=10^{-4}$ \AA/MC steps.

\begin{figure}
\vspace*{-.0cm}
\includegraphics[width=6.cm]{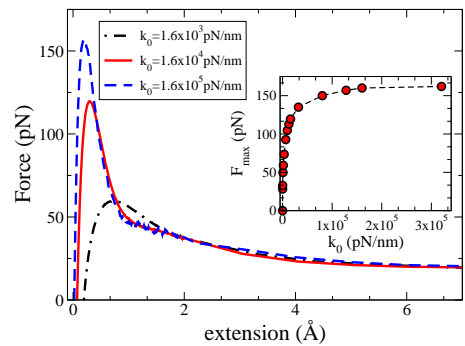}
\caption{\label{fig:barrier}
Force-extension curves of a 300 bp homogeneous AT sequence for threeÄÄ different values of the probe stiffness $k_0$. The inset shows the height of the force barrier, $F_{max}$, as a function of $k_0$.  
All curves correspond to averaging over $10^3$ Monte Carlo simulations at $T=300K$ and $v_0=0.1V$.  }
\vspace*{-.0cm}
\end{figure}

We first investigate the initiation of the unzipping process.
In Fig. \ref{fig:barrier} we present the force-extension curve of a homogeneous 300 bp AT sequence,
for different values of the stiffness, $k_0$, of the force probe.
A significant force barrier at $2$ \AA \ extensions is observed for stiff force probes.
This barrier gradually decreases with decreasing  stiffness, and finally vanishes for
$k_0 \stackrel{_<}{_\sim} 16 ~\mbox{pN/nm}$ (see inset of Fig. 2).
A simple interpretation of this result can be understood by considering the energy landscape of the entire system (DNA and force probe) \cite{shapiro}.
Assuming, for simplicity, the unzipping of the first base pair only, for small $k_0$, the total energy has two local minima separated by an activation barrier. 
The first minimum corresponds to the bound state of the base pair and the second to the unbound state. As the force probe moves, the energy landscape is 
increasingly tilted, and eventually the second minimum becomes the global minimum.  Beyond this point the unbound state is the energetically most favorable. 
In the simulation (and also in the experiment) only the transition between the bound and unbound state can be observed. States close
to the activation barrier are difficult to probe, since the probability decays exponentially with increasing barrier height.
For large $k_0$, however, the elastic energy of the force sensor becomes dominant and the total energy has only a single minimum such that the unzipping process 
is continuous with no inaccessible regions. In fact, for very stiff probes ($>1.6 \times 10^5$ pN/nm) the resulting force accurately represents the derivative of the 
potential energy with  respect to the extension. In the PBD model, the force barrier originates from a combination of the force needed to break the hydrogen bond  and the force needed to 
overcome the entropic barrier of the stacking interaction \cite{cocco,indians}.
However, it is important to notice that this barrier can be observed only for the unzipping of the first
base pairs.  When the first base pair is unzipped the effective probe consisting of the actual force probe 
and the newly formed single strand becomes very soft and the observation of the force barrier is no longer possible. 
An experimental indication
of the existence of this force barrier can be found in the work of Krautbauer {\em et al.} \cite{unzipoligo}, where 
significant force barriers are observed at the initiation of the unzipping process. However these authors attributed this barrier to interactions
with other molecules.  
The experimental ambiguity of this barrier may be attributed to two factors: First, the experimental apparatus does not have
a resolution of $2$ \AA~ where the barrier exists. Second,
the typical total stiffness of the force probe and the polymeric linkage is very small. 
The question of whether this force barrier exists when enzymes bind the DNA molecules obviously
depends on how stiff their interaction is. 
The development of more sophisticated experimental techniques able to accurately probe protein-DNA interaction, will shed light on to what extend the binding process is driven by enzymes or  thermal fluctuations \cite{kimtra} and supercoiling effects \cite{Benham}.
In what follows, we use $k_0=1.6$ pN/nm, corresponding to the value of the force probe stiffness between
the lower limit of AFM cantilevers and the upper limit of a typical optical tweezer.

\begin{figure}[h]
\vspace*{-0.cm}
\includegraphics[width=6.5cm]{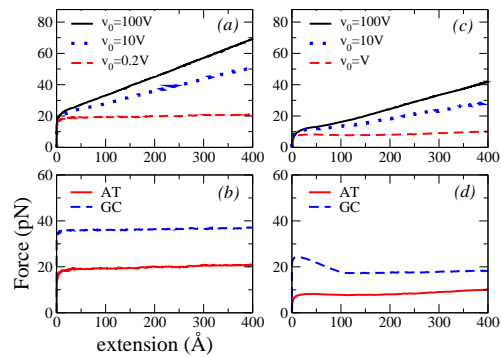}
\caption{\label{fig:fj}
Force-extension curves obtained through Monte Carlo simulations for
(a) AT sequence at three different pulling speeds, 
(b) AT and  GC sequences at $v_0=0.2V$ and $v_0=0.1V$, respectively.
Subfigures (c) and (d) are the same as (a) and (b) but using the improved description of the singe strand stacking interaction (see text for details). In (d) $v_0=V$ and $v_0=0.16V$ for AT and  GC sequences, respectively.
Each curve is the average over $10^3$ independent simulations.
In all cases $T=300K$, $k_0=1.6$ ~pN/nm and the number of unzipped base pairs is 300.}
\vspace*{-0.cm}
\end{figure}

One of the most common questions investigated by DFS experiments is the dependence of the unzipping force on pulling speed.
In Fig. \ref{fig:fj}(a) we present the force-extension curve of a homogeneous AT sequence for three different values of $v_0$, at $T=300K$. 
It is seen that faster pulling leads to higher measured force, in agreement with experiments \cite{evans99}.
However, for small $v_0$ ($\sim 0.2V$) the measured force remains practically constant during the simulation, indicating that
the system is close to equilibrium. This force corresponds to the experimentally observed unzipping force.
During the unzipping of long DNA molecules, a slight increase in the slope of the 
force-extension curve is also observed, corresponding to the elastic energy of the stretched single strand and force probe system \cite{heslot}.
It should also be noted that for even  slower pulling speeds the molecule unzipping can take place more easily. This occurs because
there is sufficient time for the system to be  
stochastically driven over the activation barrier at even lower forces \cite{evans99}.

According to our numerical results, presented in Fig. \ref{fig:fj}(b), the unzipping force of a homogeneous  AT and GC sequence is 
$20$ and $36$ pN, respectively. These values are roughly twice what one would expect from experiments \cite{HeslotATGC, RiefATGC}.
 To understand the source of this difference  one should recall that the PBD model was originally developed in order to study  
unforced thermal denaturation. In this context, the choice
 of only one degree of freedom, which describes the relative motion of a base pair, was sufficient.
 However, this picture does not accurately describe the single strand dynamics in the kind of experiments we are considering here. 
During unzipping, two dynamically uncorrelated single strands form, and so at the macroscopic scale of the experiment
there is a small possibility for two long single strands to be completely recombined. 
In the PBD model, the single strands are always dynamically correlated. As a result, in our numerical simulations, the single strands lack entropy, 
resulting in artificially high unzipping forces.
In reality, $j$ unzipped base pairs produce two single strands each consisting of $j$ bases. 
 The two single strands should be considered as a series of springs with effective stiffness  $k/2j$, rather than as in the PBD model where 
the effective stiffness is  $k/j$.
 A simple way of imposing this effective entropy reduction
on the PBD model is to reduce the strength of the stacking interaction, in the single-stranded region, to $k/2$.   
This requires the introduction  of a displacement threshold $y_{th}$ beyond which this transition takes place. 
The choice of this threshold does not critically affect the result, since
our interest lies in describing the problem at the  macroscopic scale of the  experiments.   

In Fig. \ref{fig:fj} (c) and (d) we present the numerical results obtained from this approach. The resulting rate dependence on the calculated 
force is exactly the same as in Fig. \ref{fig:fj}(a),
but the unzipping force of homogeneous AT and GC sequences is $9$ pN and $17.5$ pN, respectively, values that are in excellent agreement with
the  existing experimental results \cite{HeslotATGC, RiefATGC}. 
It is important to emphasize that {\em in vivo} unzipping is expected to require more force than the experiments indicate \cite{steveBio}.
Existence of the force barrier and the fact that the unzipping experiments probe the locally required force through a long 
and flexible single strand linker yields a lower unzipping force.
In nature, enzymes apply the force directly to the DNA lacking the flexible linker that significantly lower the measured 
force \cite{evans99}. Indeed, simulations of the unmodified PBD model indicate that an enzyme is required to apply a force of at least $20-36$ pN to unzip 
the DNA molecule. It is clear that our model and numerical approach will be able to shed new light on fundamental processes, such 
as replication and transcription when applied to interpret single-molecule exprimental data obtain for these process \cite{helic, RNAp}. 
\begin{figure}[t]
\vspace*{-0.cm}
\includegraphics[width=6.cm]{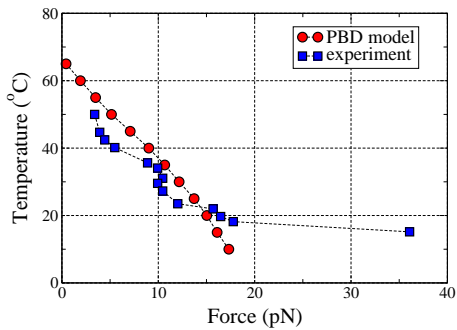}
\caption{\label{fig:pd}
Force-temperature phase diagram  the $\lambda$-phage DNA. Squares correspond to the experimental results of Ref. \cite{pd_exp}. Circles represents our numerical results for $k_0=1.6$ pN/nm and $v_0=0.16V$. Each point is the mean value of the force-extension curve obtained by unzipping the first 300 base-pairs of the molecule for  $10^3$ different simulations.}
\vspace*{-0.cm}
\end{figure}

With the  Monte Carlo technique  and the single strand modification to the PBD model described earlier, we are now able to determine the force required to unzip DNA at 
a given temperature and compare directly with the experimental results of Danilowicz {\em et al.} \cite{pd_exp}. These authors presented the force-temperature 
phase diagram of the $\lambda$-phage DNA at a constant force, compared with the prediction of a simple thermodynamic model.
Figure \ref{fig:pd} depicts our results obtained for the unzipping of the first $300$ bp
of the $\lambda$-DNA.
Since the molecule under investigation is not homogeneous, the unzipping force at each temperature corresponds to the mean value of the force-extension curve.
As  shown, our results are in very good agreement in the temperature range of $20 - 50~^o$C. We notice
a difference between our prediction and the experimental results for $T<20~^o$C. 
This low temperature behavior  may be due to a configurational change \cite{pd_exp}, or to time scales of experimental techniques.
Our predictions are more consistent with the experimental results 
than those of the thermodynamic model. This improvement is most evident at high temperatures, with the thermodynamic model yielding significantly higher forces. This is mainly 
attributed to the fact that the spontaneous bubble formation for $T>40~^o$C \cite{Zocchi}, is not described by the thermodynamic model. The bubble formation, which is predicted by the
 PBD model \cite{ares}, leads to a considerable reduction of the measured force at those higher temperatures.

In summary we have presented a  Monte Carlo approach to mechanical unzipping of DNA, within the framework of the simple PBD model. Our method allows us 
to investigate the dependence on unzipping speed and temperature. We demostrated and analyzed the existence of a force barrier at the initiation of forced unzipping which has 
generally been ignored in experimental setups. Further, we found that the single strand, as it extends between the double stranded molecule and 
the force probe, causes a decrease in the measured force. We showed that a modification of the PBD model is necessary to appropriately account for 
this effect and match experimental results. However, we find that the unmodified PBD model reliably captures the local dynamics involved in the unzipping of DNA.  With the described  modification, we were able to quantitatively  reproduce the experimental force-temperature phase diagram recently obtained for  $\lambda$-phage DNA. Evidently, the PBD model successfully encompass both the traditional thermal
separation and the more recently investigated force-induced DNA unzipping, and therefore offers significant predictive power for {\em in vivo} situations as 
well as emerging technologies.

We  thank Dr. Steven Koch for informative discussions.  Work at Los Alamos is performed
under the auspices of the US Department of Energy under contract W-7405-ENG-36.


\end{document}